\documentclass[aps,pra,amssymb, amsmath,nobibnotes,nobibnotes,reprint,superscriptaddress,showpacs,showkeys]{revtex4-1}
\usepackage{graphics}
\usepackage{epstopdf}
\usepackage{graphicx,graphics,color,epsfig}
\usepackage{revsymb4-1}
\usepackage{bm}
\usepackage{braket}
\usepackage[colorlinks, linkcolor=blue, anchorcolor=blue, citecolor=blue]{hyperref}

\begin{document}

\title{Ground-State Cooling in Cavity Optomechanics with Unresolved Sidebands}

\author{Saud Al-Awfi}
\affiliation{Department of Physics, Faculty of Science, Taibah University,
P.O.Box 30002, Madinah, Saudi Arabia}
\author{Mohannad Al-Hmoud }
\author{Smail Bougouffa}
\email{sbougouffa@hotmail.com and sbougouffa@imamu.edu.sa}
\affiliation{Physics Department, College of Science, Al Imam Mohammad ibn Saud Islamic University (IMSIU), P.O. Box 90950, Riyadh 11623, Saudi Arabia.}

\date{\today}

\begin{abstract}
We consider a simple cavity optomechanics and study the ground-state cooling of mechanical resonator in the quantum regime.  Using the effective master equations in the linear regime, the equations of motion can be obtained for the second order moments. The steady state solutions are derived in the case where the antiresonant terms are ignored. The final mean value of phonon number is compared the case where the antiresonant terms are included. We find that the ground-state cooling in the last case is improved. Indeed, the inclusion of the antiresonant terms makes the system able to generate a squeezed field, which is required for enhancing cooling. The variances of the resultant field are presented. Analytic calculations are presented in some appropriate regimes. Then our analytic predictions are confirmed with numerical calculations.
\end{abstract}

\pacs{42.50.Lc, 42.50.Wk, 07.10.Cm, 42.50.Ct.}

\maketitle

\section{Introduction}
The radiation-pressure interaction between mechanical degrees of freedom and the modes of the electromagnetic field inside an optical or microwave cavity has investigated, in order to make the cavity optomechanics more promising domain of research . The cavity optomechanics \cite{1,2,02,3,4,5,6,7} offers an interesting framework for diverse applications, such as the ultrasensitive measurements \cite{8,9}, transducing quantum communication between diverse parts of quantum networks \cite{10, 11}, testing quantum mechanics in various microscopic scales \cite{12, 13} and quantum information processing \cite{11,14}.\\
More recently, hybrid optomechanical systems have been exploited, in order to apply the features of various quantum systems to original quantum technologies \cite{15}. Particularly,  hybrid optomechanical systems have been explored  for macroscopic ground-state cooling \cite{16,17,18,19}, optomechanical coupling enhancement \cite{20}  and  entanglement \cite{21, 22, 23}.\\
With the current developements in laser cooling techniques \cite{11}, construction of low-loss optical mechanism and high- Q mechanical resonators, it is now attainable to set up nanomechanical oscillators, which may be controlled to a very high precision and can still achieve the quantum regime of the oscillations. Recently, the quantum cooling of macroscopic mechanical resonator has been proved theoretically and achieved experimentally \cite{24}.  The current results suggest that the generation of quantum resonators with a mass at the microgram scale is within reach \cite{24}. For optomechanical systems, cavity-assisted backaction cooling of mechanical resonators, cooling in the single-photon strong regime of cavity optomechanics, single-photon optomechanics, sideband cooling beyond the quantum backaction limit with squeezed light \cite{25,26} and ground-state cooling of mechanical resonators \cite{27,28,29} have been studied.\\
In this paper, we explore the effect of the unresolved sidebands of the ground-state cooling of mechanical resonators using a simple cavity optomechanics. Using the master equation in the linearized approximation, the equations of motion of the second order moments are derived. A comparison between the steady-state solutions of the obtained system with and without antiresonant terms is presented. The mean value of the phonon number is calculated in both situations. It has shown that the inclusion of the antiresonant terms improve the ground-state cooling of mechanical resonator. A simple interpretation is given, i.e., the inclusion of the antiresonant terms leads the system itself to generate a squeezed field, which is required to enhance the cooling of mechanical resonators.
The variance of the combined optical and phonon fields are calculated in the sideband regions and it has shown that the the resultant field can exhibit a squeezing of high order. For more realistic situations, we consider the situation within the recent experimental range of parameters.\\
This paper is organized as follows: In Sec. II, we introduce the model and we deduce the equations of motion for the second order moments in linearized approximation.
In Sec. III, we investigate the steady state solutions with and without the counter-rotating terms. The mean values of the steady state phonon number are obtained and compared to each other.  Some analytical and numerical solutions are presented. In Sec. IV, we present analytical and numerical calculations of the variances of the combined fields in the cavity optomechanics. We show that the resultant fields in the cavity can be squeezed in the resolved sideband cooling. In Sec. V, we summarize our remarks.

\section{Model, Hamiltonian and master equation}
In this section, an optical Fabry-Perot cavity is considered such that one mirror is driven by laser and the other mirror is able to move by the effect of the radiation pressure force (Fig.~\ref{Fig01}).
\begin{figure}[ht]
 \centerline{\includegraphics[width=0.60\linewidth,height=0.4\linewidth]{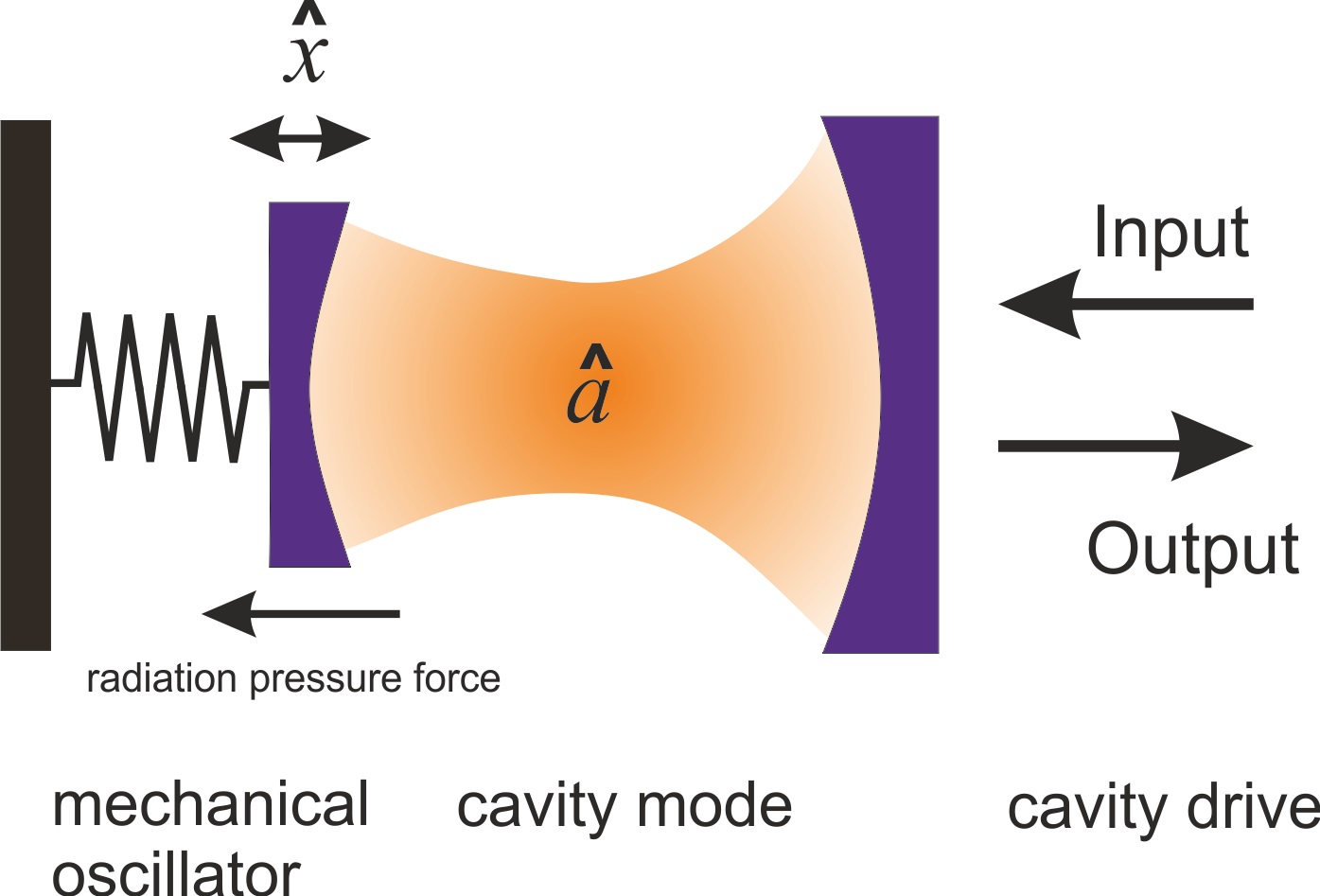}}
\caption{(Color online) Schematic of optomechanical system: Laser-driven optical cavity and a vibrating end mirror.}\label{Fig1}
\end{figure}

In order to simplify the mathematical treatment we consider one cavity mode case and we assume that the photon
scattering into other modes is ignored. As the cavity is driven by external
field, thus a quantum mechanical description of a cavity is given by the
input-output theory. On the other hand, the mechanical resonator is quantized,
thus it can be considered as an ensemble of phonon modes that are
characterized by the frequency $\omega_{m}$. In the adiabatic limit, the system Hamiltonian reads \cite{29}
\begin{eqnarray}\label{Eq1}
\hat{H}= && \hbar \omega_{c}\hat{a}^\dagger \hat{a} + \hbar \omega_{m}\hat{b}^\dagger \hat{b}
- \hbar g_{0}\hat{a}^\dagger \hat{a}(\hat{b}^\dagger+\hat{b})\nonumber\\
&+& i\hbar (E e^{-i\omega_{l}t}\hat{a}^\dagger + E^* e^{i\omega_{l}t}\hat{a}),
\end{eqnarray}
where $\omega_c$ is the optical angular resonance frequency, $\hat{a}^\dagger$ ($\hat{a}$) and $\hat{b}^\dagger$ ($\hat{b}$) are the creation (annihilation) operators of the cavity and the mechanical mode, respectively. The third term describes the optomechanical interaction with the coupling strength $g_{0}$. The last term in Eq.(\ref{Eq1}) illustrates the driving field with frequency $\omega_{l}$ and $E$ denotes the driving strength.

An appropriate investigation of the problem requires including different
effects. The main effect is the photon losses in the optical cavity that is characterized by the decay rate $\kappa
$ and the loss of mechanical excitations, i.e., phonons, which is quantified
by the energy dissipation rate $\gamma_m$.  The motion's equations can be obtained using
Heisenberg equation, thus the subsequent coupled system of
nonlinear Langevin equations\cite{30} read
\begin{eqnarray}\label{Eq3}
\dot{\hat{a}}&=&-(\frac{\kappa}{2}+i \Delta_{0})\hat{a} +i g_{o} \hat{a}(\hat{b}+\hat{b}^{\dag}) + E +\sqrt{2\kappa}\hat{a}^{in}, \nonumber\\
\dot{\hat{b}}&=&-i\omega_{m}\hat{b} -\frac{\gamma_{m}}{2} \ \hat{b} + i g_{o} \hat{a}^{\dag}\hat{a}+ \sqrt{2\gamma_m}\hat{b}^{in},
\end{eqnarray}
where $ \Delta_{0} = \omega_{c} - \omega_{l}$ is the cavity detuning,
$\kappa$ is the damping rate of the cavity mode and $(a^{in}(t))$ is the
annihilation operator of the input field, verifying the correlation relation \cite{31}
\begin{eqnarray}\label{Eq4}
\langle \hat{a}^{in}(t)\hat{a}^{in^{\dag}}(t')\rangle&=\big(N(\omega_c)+1\big)\delta(t-t') \nonumber\\
\langle \hat{a}^{in^{\dag}}(t)\hat{a}^{in}(t')\rangle&=N(\omega_c)\delta(t-t'),
\end{eqnarray}
where $N(\omega_c)=\big(\exp(\hbar \omega_c/k_B T)-1 \big)^{-1}$ is the equilibrium mean thermal photon number and $T$ is the temperature of the reservoir and $k_B$ is the Boltzmann constant. For optical frequencies $\hbar \omega_c/k_B T \gg1$, which yield $N(\omega_c)\simeq 0$, thus the only correlation function of first equation in Eq.(\ref{Eq4}) is significant.
The vibrational mode is affected by a damping force of decay rate
$\gamma_{m}$ and  the mechanical noise where the hermitian Brownian noise operator $\hat{b}^{in}$ does not describe a Markovian process. On the other hand, the quantum effects are reachable just by means of vibrations with a large mechanical quality factor $Q_{m_{j}}=\omega_m/\gamma_m \gg 1$. In this limit, we recover a Markovian process and the following second moments \cite{32}
\begin{equation}
\langle \hat{b}^{in}(t)\hat{b}^{in^{\dag}}(t')\rangle \simeq\ (\bar{n}+1)\delta(t-t'),
\end{equation}
where $\bar{n} = (e^{\hbar \omega_{m}/k_{B} T} - 1)^{-1} $ is the mean thermal excitation number at the frequency of the mechanical mode.\\
In order to explore the cooling mechanism, we use a canonical transformation of the type $\hat{a} = a_{s}+\delta \hat{a} , \hat{b}= b_{s}+\delta \hat{b}$, where the amplitudes $a_{s}, b_{s}$ are the steady state solutions of equations (\ref{Eq3}). For small parameters $g_0$ and $1/Q_m$ we get $b_{s}=g_0|a_{s}|^2/\omega_m$ and $a_s=E/(\kappa/2 + i\Delta)$, where $\Delta=\Delta_0-g_0^2|a_s|^2/\omega_m$ is the effective detuning, which includes the radiation pressure effects. Thus, $|a_s|^2$ is the steady state occupancy of the cavity in the absence of optomechanical coupling and $b_s$ is the static shift of the mechanical amplitude due to radiation pressure. In the parameter regime $|a_s|\gg 1$, one can carefully ignore the nonlinear terms $\delta \hat{a} \delta \hat{a}^\dag$ and $\delta \hat{b}\delta \hat{a}$. Then we get a system of linearized quantum Langevin equations
\begin{eqnarray}\label{Eq5}
\delta \dot{\hat{a}}&=&-(\frac{\kappa}{2}+i \Delta)\delta\hat{a} +i g_{o}a_s(\delta\hat{b}+\delta\hat{b}^{\dag})+\sqrt{2\kappa}\hat{a}^{in}, \nonumber\\
\delta\dot{\hat{b}}&=&-(\frac{\gamma_m}{2} +i\omega_{m})\delta\hat{b} + i g_{o}( a_s^{*}\delta\hat{a}+a_s\delta\hat{a}^{\dag})+ \sqrt{2\gamma_m}\hat{b}^{in}.\nonumber \\
\end{eqnarray}
Within this transformation, the corresponding Hamiltonian reads
\begin{equation}\label{Eq6}
\hat{H}_{lin}=-\hbar \Delta \delta\hat{a}^\dagger \delta\hat{a} + \hbar \omega_{m}\delta \hat{b}^\dagger \delta\hat{b}
- \hbar( g \delta\hat{a}^\dagger+ g^*\delta\hat{a})(\delta\hat{b}^\dagger+\delta\hat{b}),
\end{equation}
where $g=g_0 a_s$ illustrates the light-enhanced optomechanical coupling strength. Without lost of generality,
the reference point of the cavity field is selected such that the mean value $a_{s}$  must be real positive. In addition, the rotating terms in this Hamiltonian describe the beam splitter interaction although the counter rotating terms report the two-mode squeezed interaction. On the other hand, the operators $\delta \hat{a}$ and $\delta \hat{b}$ defining the fluctuations about the steady state values $a_s$ and $b_s$, respectively, favour equations of motion identical to the quantum master equation
\begin{eqnarray}\label{Eq7}
\dot{\rho}=&-&\frac{i}{\hbar}\big[\hat{H}_{lin},\rho\big]+\frac{\kappa}{2}\mathcal{D}[\delta \hat{a}]\rho +\frac{\gamma_m}{2}(\bar{n}+1)\mathcal{D}[\delta \hat{b}]\rho \nonumber \\ &+&\frac{\gamma_m}{2}\bar{n}\mathcal{D}[\delta \hat{b}^\dag]\rho,
\end{eqnarray}
where $\mathcal{D}[\hat{o}]\rho=\big[\hat{o} \rho,\hat{o}^\dag\big]+\big[\hat{o}, \rho \hat{o}^\dag\big]$ is the standard dissipator in Lindblad form, which remains invariant under the previous canonical transformation.\\
Using the master equation (\ref{Eq7}), the evolution of the mean phonon number $\bar{N}_b=\big<\delta \hat{b}^\dag \delta \hat{b}\big>= Tr(\rho \delta \hat{b}^\dag \delta \hat{b})$ can be given by a linear system of coupled ordinary differential equations relating all the independent second order moments
\begin{eqnarray}\label{Eq8}
\bm{\mu}=&&{}\big(\bar{N}_a,\bar{N}_b, \big<\delta \hat{a}^\dag \delta\hat{b}\big>, \big<\delta \hat{a} \delta\hat{b}^\dag\big>, \big<\delta \hat{a}\delta\hat{b}\big>, \big<\delta \hat{a}^\dag \delta\hat{b}^\dag\big>,\nonumber \\ &&\big<\delta \hat{a}^2\big>, \big<{\delta \hat{a}^\dag}^2 \big>, \big<\delta\hat{b}^2\big>, \big<{\delta \hat{b}^\dag}^2\big> \big)^T,
\end{eqnarray}
which determine the covariance matrix. The system can be read as

\begin{widetext}
\begin{eqnarray}\label{Sys9}
\frac{d}{dt}\big< \delta \hat{a}^{\dag} \delta \hat{a} \big>&=&-\kappa \big< \delta \hat{a}^{\dag} \delta \hat{a}\big> +ig\big<(\delta \hat{a}^{\dag}- \delta \hat{a})( \delta \hat{b}^{\dag}+ \delta \hat{b}) \big>, \nonumber \\
\frac{d}{dt}\big< \delta \hat{b}^{\dag} \delta \hat{b} \big>&=&-\gamma_m\big< \delta \hat{b}^{\dag} \delta \hat{b}\big>
+\gamma_{m} \bar{n}+ig\big<(\delta \hat{a}^{\dag}+ \delta \hat{a})( \delta \hat{b}^{\dag}- \delta \hat{b}) \big>, \nonumber \\
\frac{d}{dt}\big< \delta \hat{a} \delta \hat{b}^{\dag} \big>&=&-\big(\frac{\kappa+\gamma_m}{2}-i\Delta-i\omega_{m}\big)\big< \delta \hat{a} \delta \hat{b}^{\dag} \big> +ig\big(\big< \delta \hat{b}^{\dag} \delta \hat{b} \big>-\big< \delta \hat{a}^{\dag} \delta \hat{a} \big>+\big< {\delta \hat{b}^{\dag}}^{2}\big>-\big< \delta \hat{a}^{2}\big>\big),\nonumber \\
\frac{d}{dt}\big< \delta \hat{a}^{\dag} \delta \hat{b} \big>&=&-\big(\frac{\kappa+\gamma_m}{2}+i\Delta+i\omega_{m}\big)\big< \delta \hat{a}^{\dag} \delta \hat{b} \big> -ig\big(\big< \delta \hat{b}^{\dag} \delta \hat{b} \big>-\big< \delta \hat{a}^{\dag} \delta \hat{a} \big>+\big< \delta \hat{b}^{2}\big>-\big< {\delta \hat{a}^{\dag}}^{2}\big>\big),\nonumber \\
\frac{d}{dt}\big< \delta \hat{a} \delta \hat{b} \big>&=&-\big(\frac{\kappa+\gamma_m}{2}-i\Delta+i\omega_{m} \big)\big< \delta \hat{a} \delta \hat{b} \big> +ig\big(1+\big< \delta \hat{b}^{\dag} \delta \hat{b} \big>+\big< \delta \hat{a}^{\dag} \delta \hat{a} \big>+\big< \delta \hat{b}^{2}\big>+\big< \delta \hat{a}^{2}\big>\big), \nonumber \\
\frac{d}{dt}\big< \delta \hat{a}^{\dag} \delta \hat{b}^{\dag} \big>&=&-\big(\frac{\kappa+\gamma_m}{2}+i\Delta-i\omega_{m} \big)\big< \delta \hat{a}^{\dag} \delta \hat{b}^{\dag} \big> -ig\big(1 +\big< \delta \hat{b}^{\dag} \delta \hat{b} \big>+\big< \delta \hat{a}^{\dag} \delta \hat{a} \big>+\big< {\delta \hat{b}^{\dag}}^{2}\big>+\big< {\delta \hat{a}^{\dag}}^{2}\big>\big), \nonumber \\
\frac{d}{dt}\big< \delta \hat{b}^{2} \big>&=&-\big(\gamma_{m}+2i\omega_{m}\big)\big< \delta \hat{b}^{2} \big>
   +2ig\big< (\delta \hat{a}^{\dag}+ \delta \hat{a})\delta \hat{b}\big>, \nonumber \\
\frac{d}{dt}\big< {\delta \hat{b}^{\dag}}^{2} \big>&=&-\big(\gamma_{m}-2i\omega_{m}\big)\big< {\delta \hat{b}^{\dag}}^{2} \big>
   -2ig\big< (\delta \hat{a}^{\dag}+ \delta \hat{a})\delta \hat{b}^{\dag}\big>, \nonumber \\
\frac{d}{dt}\big< \delta \hat{a}^{2} \big>&=&-\big(\kappa-2i\Delta \big)\big< \delta \hat{a}^{2} \big>
   +2ig\big< (\delta \hat{b}^{\dag}+ \delta \hat{b} )\delta \hat{a}\big>,\nonumber \\
\frac{d}{dt}\big< {\delta \hat{a}^{\dag}}^{2} \big>&=&-\big(\kappa+2i\Delta \big)\big< {\delta \hat{a}^{\dag}}^{2} \big>
   -2ig\big< (\delta \hat{b}^{\dag}+ \delta \hat{b} )\delta \hat{a}^{\dag}\big>.
\end{eqnarray}
\end{widetext}
These equations (\ref{Sys9}) can be written in the following compact form
\begin{equation}\label{sys}
    \dot{\bm{\mu}}(t)=\bm{A \mu} + \bm{B},
\end{equation}
where $\bm{A}$ is the drift matrix, $\bm{B}$ is the vector composed of noise terms \cite{33} and $\bar{N}_a=\big<\delta \hat{a}^\dag \delta \hat{a}\big>$ is the mean photon number.

\section{Steady State Solutions}
In order to explore the stationary cooling mechanism, we need to solve the coupled system in the steady state. Indeed, we will examine the cooling effect with and without rotating-wave approximation.

\subsection{Without the Counter-Rotating Terms}
\begin{figure}[ht]
 \centerline{\includegraphics[width=0.7\linewidth,height=0.55\linewidth]{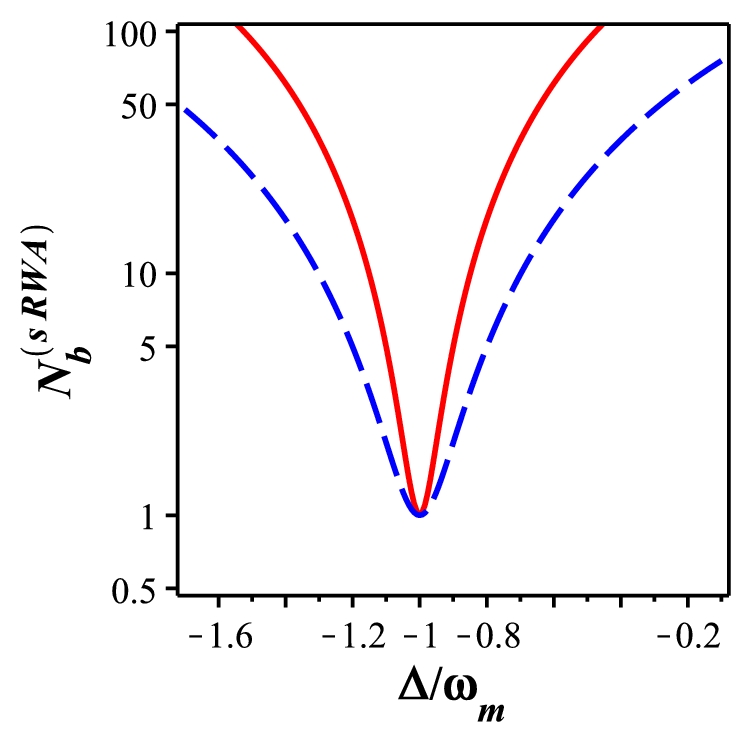}}
\caption{(Color online) Final steady state average phonon number $\bar{N}_b$ in terms of the effective normalized detuning $\Delta/\omega_m$ for different values of the ratio $g/\omega_m$  ($g/\omega_m= 0.05, 0.1$) with red solid and blue dashed lines respectively, with $\gamma_m/\omega_m=10^{-5}$, $\bar{n}=10^3$ and $\kappa/\omega_m=0.01$. }\label{Fig2}
\end{figure}

Here we are concerned with an interesting case, where $ \kappa \ll g \ll \omega_m$. In this regime, the rotating-wave approximation(RWA) can be used such that the counter-rotating terms $\delta \hat{a} \delta\hat{b}$ and $\delta \hat{a}^{\dag} \delta\hat{b}^{\dag}$ are ignored and the previous coupled system becomes
\begin{eqnarray}
\dot{\overline{N}_a}&=&-\kappa \overline{N}_a + ig\overline{C}_{-}, \nonumber \\
\dot{\overline{N}_b}&=&-\gamma_m\overline{N}_b
+\gamma_{m} \bar{n}- ig\overline{C}_{-},\nonumber \\
\dot{\overline{C}_-}&=&-\big(\frac{\kappa}{2}+\frac{\gamma_m}{2}\big)\overline{C}_{-}
-i\big(\Delta_{L}+\omega_{m}\big)\overline{C}_{+} -2ig(\overline{N}_b -\overline{N}_a),\nonumber \\
\dot{\overline{C}_{+}}&=&-\big(\frac{\kappa}{2}+\frac{\gamma_m}{2}\big)\overline{C}_{+}-i\big(\Delta_{L}+
\omega_{m}\big)\overline{C}_{-},
\end{eqnarray}
where $\overline{C}_{\pm}=\big< \delta \hat{a}^{\dag} \delta \hat{b} \big> \pm \big< \delta \hat{a} \delta \hat{b}^{\dag} \big>$ describe the coherences between the optical and mechanical modes. The steady-state solution $\bar{N}_{b}^{s, RWA}$ can be obtained as
\begin{widetext}
\begin{eqnarray}\label{Eq18}
    \overline{N}_{b}^{s,RWA}=\bar{n}\gamma_m\frac{\kappa^2(\kappa+2\gamma_m)+\kappa(\gamma_m^2+4g^2+4(\Delta+\omega)^2)+4g^2\gamma_m}
    {\gamma_m\kappa^3+(4g^2+2\gamma_m^2)\kappa^2+(\gamma_m^2+8g^2+4(\Delta+\omega)^2)\gamma_m\kappa+4g^2\gamma_m^2}.
\end{eqnarray}
\end{widetext}
For the red sideband resonant region with $\Delta=-\omega_m$, where the cooling measure is at resonance, we obtain
\begin{equation}\label{Eq19}
\overline{N}_{b}^{s, RWA}=\frac{\bar{n}\gamma_m}{\gamma_m+\kappa}\big(1+\frac{\kappa^2}{4g^2+\gamma_m \kappa} \big),
\end{equation}
which shows that the steady state cooling limit in RWA depends on the cavity and mechanical decay rate. In Fig.~\ref{Fig2} we present the variation of $\bar{N}_{b}^{s,RWA}$ in terms of the normalized effective detuning for some values of effective coupling strength.

\subsection{With the Counter-Rotating Terms}
\begin{figure}[ht]
 \centerline{\includegraphics[width=0.7\linewidth,height=0.55\linewidth]{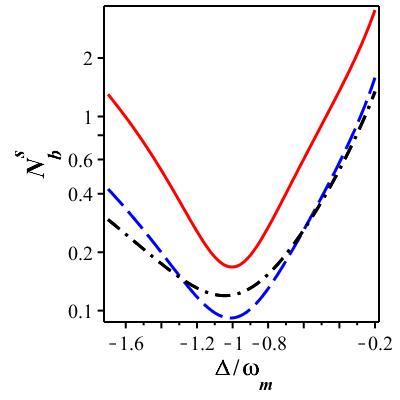}}
\caption{(Color online) Final steady state average phonon number $\bar{N}_b$ in terms of the effective normalized detuning $\Delta/\omega_m$ for different values of the ratio $g/\omega_m$  ($g/\omega_m= 0.1, 0.2, 0.3$) with red solid, blue dashed and black dash dotted lines respectively with $\gamma_m/\omega_m=10^{-5}$, $\kappa/\omega_m=0.5$ and $\bar{n}=10^3$.} \label{Fig3}
\end{figure}

On the other hand, for the regime where the coupling strength is close to mechanical resonance frequency $\omega_m$, the effect of the counter-rotating terms must be included and we have recourse to solve the complete system Eq.(\ref{sys}). As we are interested to the steady state solution $\bar{N}_{b}^s$ for the cooling mechanism, the expressions are very cumbersome, we present in Fig.~\ref{Fig3}, the variation of $\bar{N}_{b}^s$ in terms of the normalized effective detuning for different values of $g$. For other parameters, we restrict ourself to the recent experimental realizations \cite{34}.
We can see from these results that the cooling in the optomechanical cavity is enhanced by including the anti-resonance terms which describe the two-mode squeezing interaction.\\
In order to explore the analytic expressions of the final steady state average phonon number $\bar{N}_b$, we will consider the interesting regime, where $\gamma_m\ll \kappa, g, \omega_m$, which is close to the experimental realizations. In this case, we can get $(\gamma_m\rightarrow 0)$ such that $\gamma_m \bar{n}$ is kept limited. In this regime, for $\Delta=-\omega_m$ we get the minimum final average phonon number as
\begin{eqnarray}\label{Eq20}
   Min\{\overline{N}_{b}\}=&&{}\frac{\bar{n}\gamma_m}
    {64\kappa g^2 \omega_m^2}\Big( \kappa^4 + 16\omega_m^2(\kappa^2+4g^2) + 8g^2 \beta \Big) \nonumber \\& +&\frac{1}{16\omega_m^2}(\kappa^2+\beta),
\end{eqnarray}
where
\begin{eqnarray}\label{Eq21}
   \beta = \frac{8g^2(\kappa^2 +4\omega_m^2)}{\kappa^2 +4\omega_m^2 -16g^2},
\end{eqnarray}
The first term in Eq. (~\ref{Eq20}) is due to the mechanical dissipation, while the second term represents the heating generated by the quantum backaction applied by the cavity. These main contributions can be optimized in terms of the effective coupling strength $g$ and the decay rate $\kappa$. In addition, the thermal noise in the cavity input contributes to the final occupancy of the mechanical resonator, but it is canceled in this treatment by considering the mean value $N(\omega_c)\simeq0$.
\begin{figure}[ht]
 \centerline{\includegraphics[width=0.8\linewidth,height=0.6\linewidth]{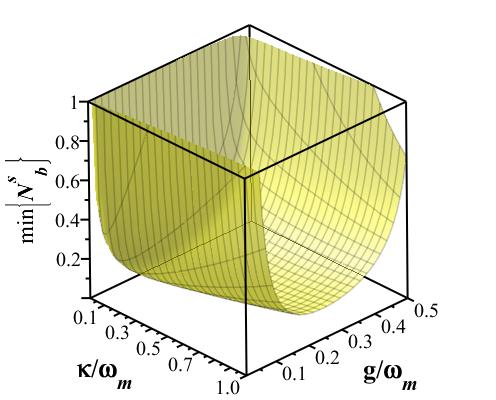}}
\caption{(Color online) Final steady state average phonon number minimized with respect to $\Delta$ in terms of  $g/\omega_m$ and $\kappa/\omega_m$ with $\gamma_m\rightarrow 0$, while $\gamma_m\bar{n}$ is kept finite.} \label{Fig5}
\end{figure}
From Fig.~\ref{Fig5}, we can understand the oscillation of $\overline{N}_{b}$ around $\Delta=-\omega_m$ in terms of the effective normalized coupling strength $g/\omega_m$. Furthermore, this result shows the regions of the parameters where the optimum final steady state average number can be obtained.\\

\section{Steady-state optomechanical squeezing}
Quantum squeezing in optomechanics systems is not only a key feature of macroscopic quantum properties \cite{25, 0035, 035}, but can also be utilized to advance the cooling of radiation-pressure interaction. In this part, we show that the robust optomechanical squeezing field in the steady state can be generated in a cavity optomechanics when the antiresonant (non-RWA) terms are included in the covariance approach for the regime where the coupling strength is close to mechanical resonance frequency $\omega_m$. The squeezing of the resultant optomechanical field is just generated by the proper incorporation of the antiresonant terms in the treatment and leads to the improvement of cooling.
\subsection{Squeezed field}
Consider a single mode field which can be written as
\begin{equation}\label{021}
    \hat{E}=E_0\big(\hat{a}e^{-i(\omega t - \vec{k}.\vec{r})}-\hat{a}^\dag e^{i(\omega t - \vec{k}.\vec{r})} \big),
\end{equation}
where $E_0$ is a constant. Introducing the two Hermitian operators $\hat{E_1}=\frac{1}{\sqrt{2}}(\hat{a}+\hat{a}^{\dag})$ and $\hat{E_2}=\frac{1}{\sqrt{2}i}(\hat{a}-\hat{a}^{\dag})$, which satisfy the commutation relation $[\hat{E_1}, \hat{E_2}]=i$. The hermitian operators $\hat{E_1}$ and $\hat{E_2}$ are completely analog to the in-phase and out-off phase quadrature components fields. The noncommutating quadrature components satisfy the Heisenberg uncertainty relation $\sqrt{\langle(\Delta \hat{E_1})^2\rangle\langle(\Delta \hat{E_2})^2}\geq \frac{1}{2}$, where $\langle(\Delta \hat{E_i})^2\rangle$ is the variance of the ith quadrature component of the field being in the state $\ket{\Psi}$. With this Heisenberg uncertainty relation, one can distinguish three basic field kinds.\\
For a chaotic field, we have $\langle(\Delta \hat{E_1})^2\rangle > \frac{1}{2}$ and $\langle(\Delta \hat{E_2})^2\rangle > \frac{1}{2}$.\\
For the vacuum or coherent field, we have $\langle(\Delta \hat{E_1})^2\rangle =\frac{1}{2}$ and $\langle(\Delta \hat{E_2})^2\rangle = \frac{1}{2}$.\\
Squeezed filed is specified by either $\langle(\Delta \hat{E_1})^2\rangle <\frac{1}{2}$ or $\langle(\Delta \hat{E_2})^2\rangle < \frac{1}{2}$.

\subsection{Variances of resultant fields}
\begin{figure}[ht]
\centering
\includegraphics[width=0.7\linewidth,height=0.6\linewidth]{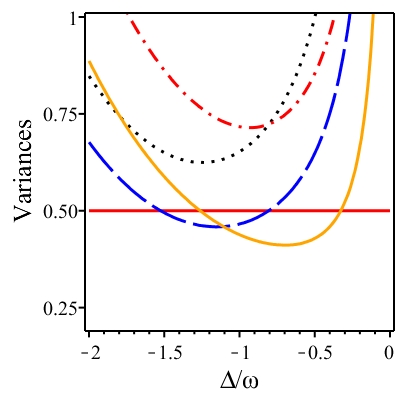}
\caption{(Color online) The variances of the quadratic components of the fields  $X_{d^{\pm}}$ and $ Y_{d^{\pm}}$ in terms of  $\Delta/\omega_m$ for $g/\omega_m=0.2$, $\kappa/\omega_m=0.5$, $\gamma_m/\omega_m=10^{-5}$ and $\bar{n}=10^3$.
The blue long dashed line for $\big<\big(\Delta X_{d^{-}}\big)^2\big>$, black dotted line $\big<\big(\Delta Y_{d^{-}}\big)^2\big>$, red dash dotted for $\big<\big(\Delta X_{d^{+}}\big)^2\big>$ and orange solid line for $\big<\big(\Delta Y_{d^{+}}\big)^2\big>$.} \label{Fig05}
\end{figure}

In the following, in the considered regime, where the anti-resonance terms are incorporated, we explore the squeezing effect of the combined optomechanical field in the considered model. Indeed, we consider the combined fields
\begin{eqnarray}\label{Eq22}
   \hat{d}^{\pm} = \frac{1}{\sqrt{2}}(\delta \hat{a}\pm \delta \hat{b}),
\end{eqnarray}
where $[ \hat{d}^{\pm}, \hat{d}^{{\pm}^{\dag}}]=1$. The quadratic components of these fields are defined as
\begin{eqnarray}\label{Eq23}
   X_{d^{\pm}} = \frac{1}{\sqrt{2}}(\hat{d}^{\pm}+\hat{d}^{{\pm}^{\dag}}),\nonumber\\
   Y_{d^{\pm}} = \frac{1}{\sqrt{2}i}(\hat{d}^{\pm}-\hat{d}^{{\pm}^{\dag}}),
\end{eqnarray}
which obey the commutation relation $[X_{d^{\pm}}, Y_{d^{\pm}}]=i$ and satisfy the Heisenberg uncertainty relation
\begin{eqnarray}\label{Eq24}
   \sqrt{\big<\big(\Delta X_{d^{\pm}}\big)^2\big>\big<\big(\Delta Y_{d^{\pm}}\big)^2\big>} \geq \frac{1}{2},
\end{eqnarray}
where the factor $"1/2"$ on the right-hand side determines the vacuum level of the fluctuations. $\big<\big(\Delta X_{d^{\pm}}\big)^2\big>$ and $\big<\big(\Delta Y_{d^{\pm}}\big)^2\big>$ are the variances of the quadrature components, which are given by
\begin{eqnarray}\label{Eq25}
   \big<\big(\Delta X_{d^{\pm}}\big)^2\big>&=\big< X_{d^{\pm}}^2\big>-\big<X_{d^{\pm}}\big>^2, \nonumber \\
   \big<\big(\Delta Y_{d^{\pm}}\big)^2\big>&=\big< Y_{d^{\pm}}^2\big>-\big<Y_{d^{\pm}}\big>^2,
\end{eqnarray}
which can be expressed in terms of the independent second order moments of Eq.~\ref{Eq7} as
\begin{widetext}
\begin{eqnarray}\label{Eq26}
\big<\big(\Delta X_{d^{+}}\big)^2\big>&=&\frac{1}{2}\Big(1+\sum_{i=1}^{6}\mu_i+ \frac{1}{2}\sum_{i=7}^{10}\mu_i\Big), \nonumber \\
\big<\big(\Delta X_{d^{-}}\big)^2\big>&=&\frac{1}{2}\Big(1+\sum_{i=1}^{2}\mu_i-\sum_{i=3}^{6}\mu_i + \frac{1}{2}\sum_{i=7}^{10}\mu_i\Big), \nonumber \\
\big<\big(\Delta Y_{d^{+}}\big)^2\big>&=&\frac{1}{2}\Big(1+\sum_{i=1}^{4}\mu_i- \sum_{i=5}^{6}\mu_i- \frac{1}{2}\sum_{i=7}^{10}\mu_i\Big),\nonumber \\
\big<\big(\Delta Y_{d^{-}}\big)^2\big>&=&\frac{1}{2}\Big(1+\sum_{i=1,2,5,6}\mu_i- \sum_{i=3}^{4}\mu_i-\frac{1}{2}\sum_{i=7}^{10}\mu_i\Big).
\end{eqnarray}
\end{widetext}
In Fig.~\ref{Fig05}, we plot the variances in terms of the normalized effective detuning $\Delta/\omega_m$ for $\kappa/\omega_m=0.5$, $g/\omega_m=0.2$, $\gamma_m=10^{-5}$ and $\bar{n}=10^3$.

\begin{figure}[h]
\resizebox{1\linewidth}{!}{%
 \includegraphics{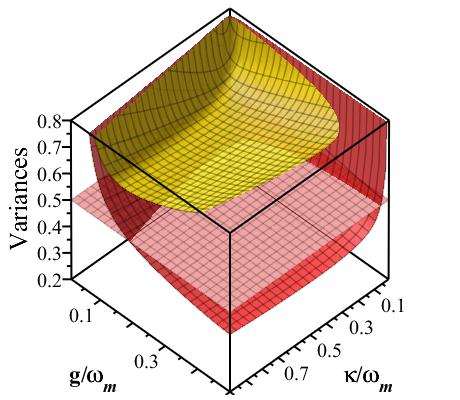}~\includegraphics{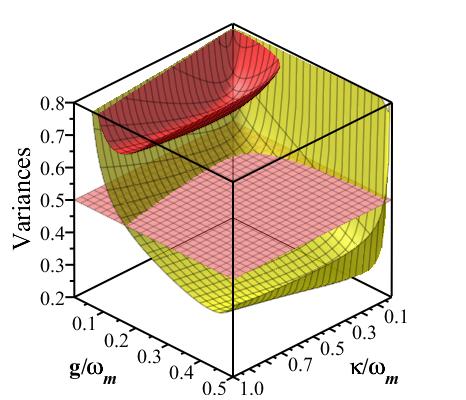}}
\caption{(Color online) The variances of the quadratic components of the fields  $X_{d^{\pm}}$ and $ Y_{d^{\pm}}$ in terms of  $g/\omega_m$ and $\kappa/\omega_m$ for $\Delta=-\omega_m$, $\gamma_m=/\omega_m=10^{-5}$ and $\bar{n}$. The left and right panels correspond to the field $d^{+}$ and $d^{-}$, respectively. The red and yellow surfaces represent $\big<\big(\Delta X_{d^{\pm}}\big)^2\big>$ and $\big<\big(\Delta Y_{d^{\pm}}\big)^2\big>$, respectively, while the orange one is the vacuum level of the fluctuations. } \label{Fig6}
\end{figure}

In the regime $\gamma_m\ll \kappa, g, \omega_m$, the analytic expressions of the variances can be obtained. The variance $\big<\big(\Delta X_{d^{+}}\big)^2\big>$ can be read as
\begin{eqnarray}\label{Eq27}
  \big<\big(\Delta Y_{d^{+}}\big)^2\big> &=& \frac{1}{2}-\big(\frac{g}{2\omega_m}-\frac{\kappa^2}{32\omega_m^2}(1+\frac{16g^2}{D})\big)+\bar{n}\gamma_m h_1,\nonumber \\
  \big<\big(\Delta X_{d^{-}}\big)^2\big> &=&{} \frac{1}{2}-\Big(\frac{g}{2\omega_m}-\frac{\kappa^2}{32\omega_m^2}(1+\frac{16g^2}{D})\nonumber \\&&-\frac{4g^2}{D}(\omega_m-2g) \Big)+\bar{n}\gamma_m h_2,
\end{eqnarray}
where
\begin{eqnarray}\label{Eq28}
h_1&=&\frac{1}{\kappa}(1-\frac{g}{\omega_m})+\frac{\kappa}{8g}(\frac{1}{g}-\frac{1}{\omega_m})\nonumber \\ {}&&+\frac{\kappa}{16\omega_m^2}(1+\frac{\kappa^2}{8g^2}+\frac{16g^2}{ D}),\nonumber \\
h_2&=&h_1+\frac{8g^2(\omega_m-2g)}{\kappa\omega_m D},
\end{eqnarray}
and
\begin{eqnarray}\label{Eq29}
D&=4\omega_m^2+\kappa^2-16g^2 ,
\end{eqnarray}
which clearly show that the variances can be less than $1/2$ for some parameter ranges. We present in Fig.~\ref{Fig6} the variances $\big<\big(\Delta X_{d^{\pm}}\big)^2\big>$ and $\big<\big(\Delta Y_{d^{\pm}}\big)^2\big>$ in terms of $g/\omega_m$ and $\kappa/\omega_m$ for $\Delta=-\omega_m$. It is obviously that the variances $\big<\big(\Delta Y_{d^{+}}\big)^2\big>$ and $\big<\big(\Delta X_{d^{-}}\big)^2\big>$ are smaller than $1/2$ in the red sideband resonant region. This means that the anti-resonance terms can generate squeezed states as well as they enhance the cooling in the optomechanical cavity. The inclusion of the anti-resonance terms make the system able to generated a squeezing light, which is required for cooling of mechanical resonators and then for quantum entanglement \cite{35}.

\section{Conclusion}
We have explored the cooling of a mechanical resonator in the quantum regime using the effective master equations for the applicable degrees of freedom in linear regime. We have derived the equations of motion for the second order moments. We have addressed the crucial point, which reside essentially in the contribution of the rotating and counter-rotating terms in the steady state solutions of the equations of motion. The final steady state mean values of the phonon number are compared in both cases. Our results showed that the incorporation of the counter -rotating terms improve the ground-state cooling of mechanical resonator in the red sideband regime, which is in good agreement with an average phonon occupation $\bar{N_b}=0.20 \pm 0.02$ that has recently been experimentally achieved \cite{36}. The analysis of the steady state of the system has demonstrated the importance of the inclusion of the antiresonant terms in the equations of motion. In addition, the insertion of the antiresonant terms leads to the creation of squeezed field in the cavity optomechanics.  We have shown that the resultant field can exhibit a squeezing of high order, which is at the heart of improving the ground state cooling. The variances of the combined optical and phonon fields are calculated and our results showed that the system can generate a squeezed field. Our results may be constructive for making distinction between different regimes of resonant and antiresonant terms, which merit a further study in the future.\\

\begin{acknowledgements}
The researchers acknowledge the deanship of Scientific Research at Al Imam Mohammad Ibn Saud Islamic University, Saudi Arabia, for financing this project under grant no. (381213)\\
We thank Z. Ficek for valuable discussions.
\end{acknowledgements}



\begin{thebibliography}{99}
\bibitem{1} M. Aspelmeyer, T. J. Kippenberg, and F. Marquardt,
Rev. Mod. Phys. \textbf{86}, 1391 (2014).

\bibitem{2} T. J. Kippenberg and K. J. Vahala, Opt. Express \textbf{15},
17172 (2007).

\bibitem{02} L.-h. Sun, G.-x. Li, and Z. Ficek, Phys. Rev. A \textbf{85},
022327 (2012).

\bibitem{3} T. J. Kippenberg and K. J. Vahala, Science \textbf{321}, 1172
(2008).

\bibitem{4} D. Rugar, R. Budakian, H. Mamin, and B. Chui, Nature
\textbf{430}, 329 (2004).

\bibitem{5} A. Schliesser and T. Kippenberg, in Cavity Optomechan-
ics (Springer, 2014) pp. 121–148.

\bibitem{6} P. Roelli, C. Galland, N. Piro, and T. J. Kippenberg,
Nat. Nanotechnol. \textbf{11}, 164 (2016).

\bibitem{7} J.-M. Pirkkalainen, S. Cho, F. Massel, J. Tuorila,
T. Heikkilä, P. Hakonen, and M. Sillanpää, Nat. Com-
mun. \textbf{6}, 6981 (2015).

\bibitem{8} J. Teufel, T. Donner, M. Castellanos-Beltran, J. Harlow,
and K. Lehnert, Nat. Nanotechnol. \textbf{4}, 820 (2009).

\bibitem{9} D. Rugar, R. Budakian, H. Mamin, and B. Chui, Nature
\textbf{430}, 329 (2004).

\bibitem{10} P. Rabl, S. J. Kolkowitz, F. Koppens, J. Harris, P. Zoller,
and M. D. Lukin, Nat. Phys. \textbf{6}, 602 (2010).

\bibitem{11} K. Stannigel, P. Komar, S. Habraken, S. Bennett, M. D.
Lukin, P. Zoller, and P. Rabl, Phys. Rev. Lett. \textbf{109},
013603 (2012).

\bibitem{12} J. Chan, T. M. Alegre, A. H. Safavi-Naeini, J. T.
Hill, A. Krause, S. Gröblacher, M. Aspelmeyer, and
O. Painter, Nature \textbf{478}, 89 (2011).

\bibitem{13} O. Romero-Isart, Phys. Rev. Lett. 107, 020405 (2011).

\bibitem{14} P. Komar, S. Bennett, K. Stannigel, S. Habraken,
P. Rabl, P. Zoller, and M. D. Lukin, Phys. Rev. A \textbf{87},
013839 (2013).

\bibitem{15} Z. Xiang, Rev. Mod. Phys. \textbf{85}, 623 (2013).

\bibitem{16} C. Genes, H. Ritsch, M. Drewsen, and A. Dantan, Phys.
Rev. A \textbf{84}, 051801 (2011).

\bibitem{17} F. Bariani, Phys. Rev. A \textbf{90}, 033838 (2014).

\bibitem{18} Z. Yi, G.-x. Li, S.-p. Wu, and Y.-p. Yang, Opt. Express
\textbf{22}, 20060 (2014).

\bibitem{19} R.-P. Zeng, S. Zhang, C.-W. Wu, W. Wu, and P.-X.
Chen, J. Opt. Soc. Am. B \textbf{32}, 2314 (2015).

\bibitem{20} H. Ian, Phys. Rev. A \textbf{78}, 013824 (2008).

\bibitem{21} C. Genes, D. Vitali, P. Tombesi, S. Gigan, and M. As-
pelmeyer, Phys. Rev. A \textbf{77}, 033804 (2008).

\bibitem{22} L. Zhou, Y. Han, J. Jing, and W. Zhang, Phys. Rev. A
\textbf{83}, 052117 (2011).

\bibitem{23} W. Ge, M. Al-Amri, H. Nha, and M. S. Zubairy, Phys.
Rev. A \textbf{88}, 022338 (2013).

\bibitem{24} T. P. Purdy, R. W. Peterson, and C. Regal, Science \textbf{339},
801 (2013).

\bibitem{25} J. B. Clark, F. Lecocq, R. W. Simmonds, J. Aumentado,
and J. D. Teufel, Nature \textbf{541}, 191 (2017).

\bibitem{26} C. Yang, L. Zhang, and W. Zhang, EPL (Europhysics
Letters) \textbf{122}, 14001 (2018).

\bibitem{27} I. Wilson-Rae, N. Nooshi, J. Dobrindt, T. J. Kippenberg,
and W. Zwerger, New J. Phys. \textbf{10}, 095007 (2008).

\bibitem{28} J. Chan, T. M. Alegre, A. H. Safavi-Naeini, J. T.
Hill, A. Krause, S. Gröblacher, M. Aspelmeyer, and
O. Painter, Nature \textbf{478}, 89 (2011).

\bibitem{29} M. R. Vanner, J. Hofer, G. D. Cole, and M. Aspelmeyer,
Nat. Commun. \textbf{4}, 2295 (2013).

\bibitem{30} D. Vitali, S. Gigan, A. Ferreira, H. Böhm, P. Tombesi,
A. Guerreiro, V. Vedral, A. Zeilinger, and M. As-
pelmeyer, Phys. Rev. Lett. \textbf{98}, 030405 (2007).

\bibitem{31} C. Genes, D. Vitali, P. Tombesi, S. Gigan, and M. As-
pelmeyer, Phys. Rev. A \textbf{77}, 033804 (2008).

\bibitem{32} V. Giovannetti and D. Vitali, Phys. Rev. A \textbf{63}, 023812
(2001).

\bibitem{33} Y.-C. Liu, Y.-F. Shen, Q. Gong, and Y.-F. Xiao, Phys.
Rev. A \textbf{89}, 053821 (2014).

\bibitem{34} A. Fainstein, N. D. Lanzillotti-Kimura, B. Jusserand,
and B. Perrin, Phys. Rev. Lett. \textbf{110}, 037403 (2013).

\bibitem{035}K. J\"{a}hne, C. Genes, K. Hammerer, M. Wallquist,
E. S. Polzik and P. Zoller P., Phys. Rev. A, \textbf{79} (2009)
063819.

\bibitem{0035} X. Y. Lü, J. Q. Liao, L. Tian and F. Nori, Phys. Rev.
A, \textbf{91} (2015) 013834.

\bibitem{35} A. A. Rehaily and S. Bougouffa, Int. J. Theor. Phys. \textbf{56},
1399 (2017).

\bibitem{36} R. Peterson, T. Purdy, N. Kampel, R. Andrews, P.-L.
Yu, K. Lehnert, and C. Regal, Phys. Rev. Lett. \textbf{116},
063601 (2016).

\end{thebibliography}

\end{document}